\documentstyle{camera}

\newcommand{\AmS}{{\protect\the\textfont2
  A\kern-.1667em\lower.5ex\hbox{M}\kern-.125emS}}

\newcommand{\beq}{\begin{equation}}
\newcommand{\eeq}{\end{equation}}
\newcommand{\bea}{\begin{eqnarray}}
\newcommand{\eea}{\end{eqnarray}}

\def\dm2{\Delta m^2}
\def\sq2{sin^2(2\Theta)}

\input epsf

\begin{document}

%
\title{HIGH ENERGY GAMMA-RAYS \\ FROM MASSIVE BINARY SYSTEMS}

%
\author{WLODEK BEDNAREK}

%
\organization{Department of Experimental Physics, University of \L \'od\'z\\ 
ul. Pomorska 149/153, 90-236 \L \'od\'z, Poland}

\maketitle

\begin{abstract}
During last years a few massive binary systems have been detected in the TeV $\gamma$-rays. This $\gamma$-ray emission is clearly modulated with the orbital periods of these binaries suggesting its origin inside the binary system. In this paper we
summarize the anisotropic IC $e^\pm$ pair cascade model as likely explanation of these observations. We consider scenarios in which particles are accelerated to relativistic energies, either due to the presence of an energetic pulsar inside the binary, or as a result of accretion process onto the compact object during which the jet is launched from the inner part of the accretion disk, or in collisions of stellar winds from the massive companions.
\end{abstract}
\vspace{1.0cm}
\section{Introduction}

All detected binary systems at TeV $\gamma$-rays belong to the class of high 
mass binaries in which compact object (neutron star or black hole) appears very close to the stellar surface at least during a part of its orbit. In fact, TeV $\gamma$-rays have been expected on theoretical grounds from the well known Be binary system, PSR B1259-63/SS2883, containing 47.7 ms pulsar since it was well known that even isolated pulsars with similar parameters should emit TeV $\gamma$-rays (e.g. the Crab pulsar and its nebula). Such emission has been discovered from this object during the periastron passage of the pulsar in 2004 (Aharonian et al.~2005a).
A year later, emission from two compact binary systems, LS 5039 and LSI +61 303, classified as microquasars,  have been observed by HESS and MAGIC telescopes (Aharonian et al.~2005b, Albert et al.~2006). This emission is clearly modulated with the orbital periods of the binary systems (Aharonian et al.~2006, Albert et al.~2008).
Moreover, there are evidences of detection of the TeV $\gamma$-ray flare from another microquasar Cyg X-1 (Albert et al.~2007).
Finally, the HESS discovered the TeV emission from the open cluster Wester\-lund 2 which contains the most massive binary system of two WR type stars, WR 20a
(Aharonian et al.~2007). This emission might also be related to this very unusual binary system.

TeV $\gamma$-ray production in binary system containing energetic pulsar is usually interpreted in terms of the leptonic IC scattering model in which relativistic electrons scatter soft radiation coming from the massive star (e.g. see scenarios discussed by Maraschi \& Treves~1981, Tavani \& Arons~1997, Kirk et al.~1999).  Similar radiation processes have been also considered in more detail in the case of electrons accelerated in the jets of  microquasars (e.g. Bosch Ramon et al.  2006, Dermer \& B\"ottcher~2006).
In these calculations the absorption of $\gamma$-ray photons on the stellar radiation is usually not taken into account or considered only in approximate way.
More recently $\gamma$-ray emission from one-directional IC $e^\pm$ pair cascades 
(secondary leptons move in the direction of primary $\gamma$-ray photons)  have been considered in a few papers (Aharonian et al. ~2006, Orellana et al.~2007, Khangulyan 
et al.~2008).
 
Here, we discuss the three dimensional IC $e^\pm$ pair cascade scenario for the TeV $\gamma$-ray origin in the massive binary systems which have been 
developed in the late 90-ties (e.g. Bednarek~2000) and later applied to these type of sources in a sequence of recent papers (Bednarek~2005, 2006a,b, 2007, Bednarek \& Giovannelli~2007, Sierpowska-Bartosik~2007, Sierpowska \& Bednarek~2005,  Sierpowska-Bartosik \& Bednarek~2008, Sierpowska-Bartosik \& Torres~2007, 2008).

\section{Optical depths for $\gamma$-rays}

\begin{figure*}[t]
\vskip 6.truecm
\includegraphics{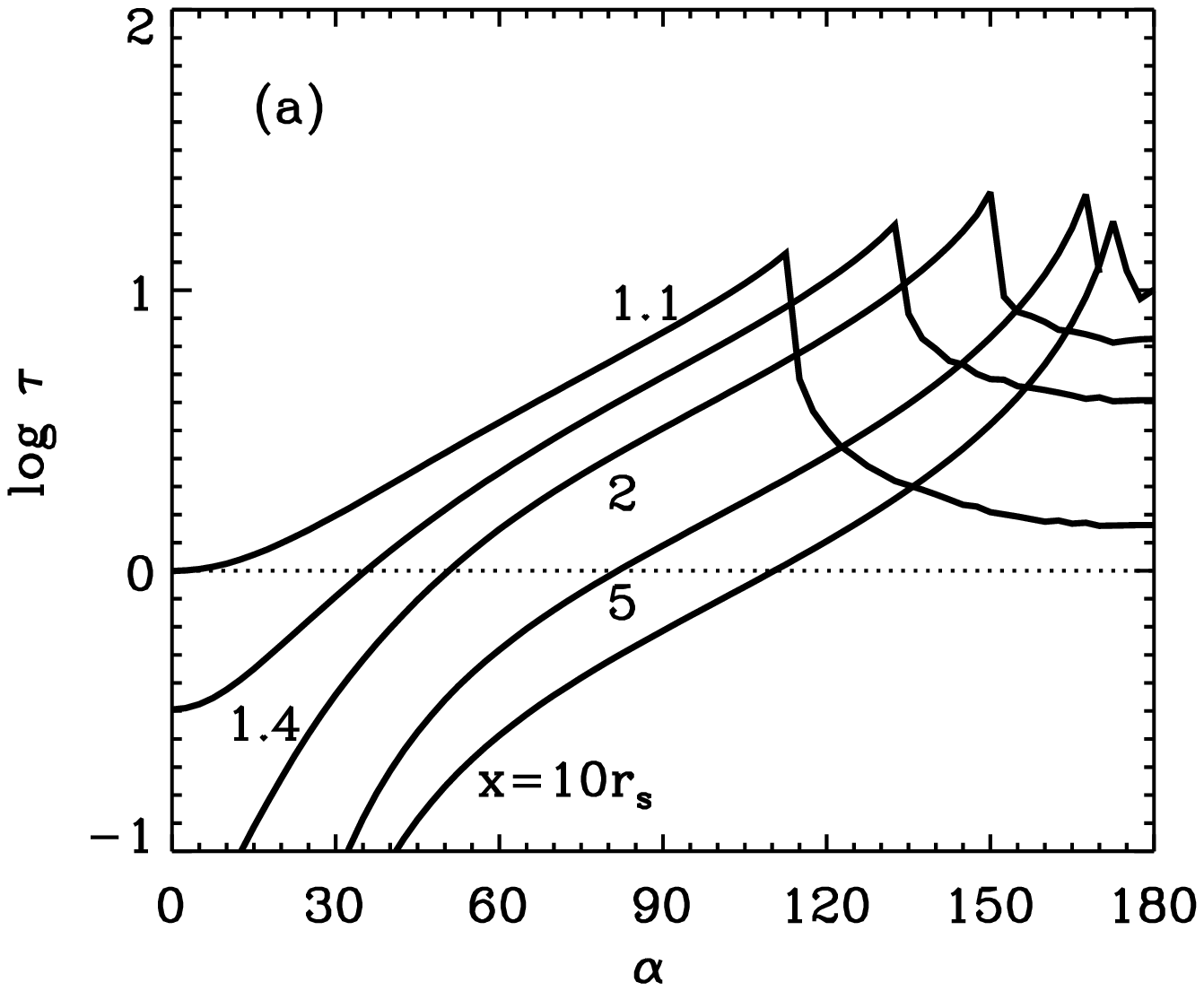}
\includegraphics{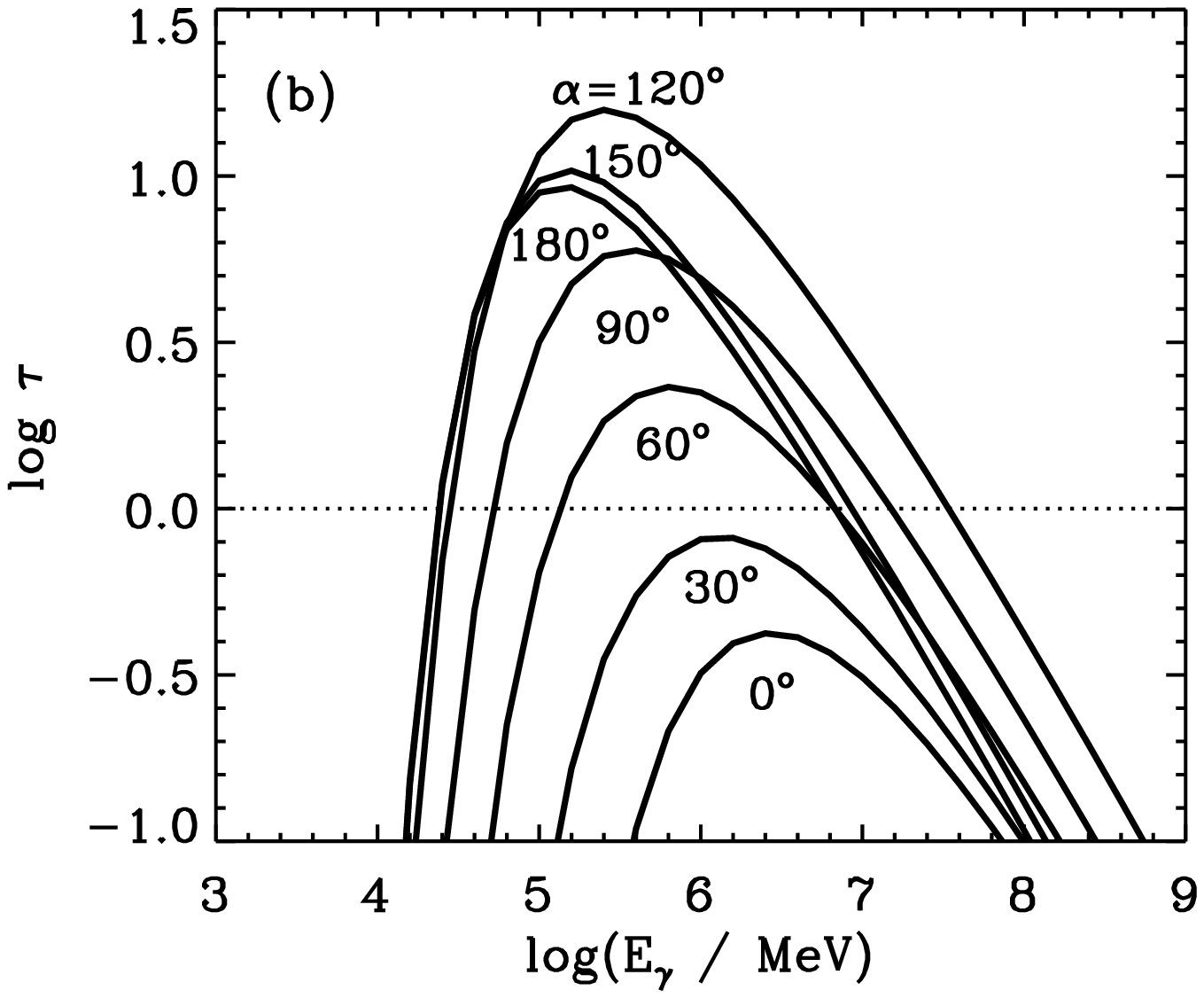}
\caption{The optical depths for $\gamma$-rays in the typical radiation field of the massive star
with the radius $R^\star = 8.6\times 10^{11}$ cm, and surface temperature $T^\star = 3\times 10^4$ K (Cen X-3 binary system), from Bednarek (2000). (a) Dependence of the optical depths on the injection angle measured from the direction defined by the injection place and the center of the star ($\alpha = 0^o$ the outward direction from the star), for selected distances of the injection place 
(in units of $R^\star$), and for the $\gamma$-ray energy  $E_\gamma = 1$ TeV. (b) Dependence on the $\gamma$-ray energy for fixed injection distance  $D = 1.4R^\star$. } 
\label{fig1}
\end{figure*}

The optical depth in the massive star radiation has been calculated in a few papers
with different simplified assumptions. The most general one (which can be applied to the 
case of injection source at arbitrary location around the massive star) can be found in
Bednarek (2000). These calculations take into account the dimensions of the massive 
star which is specially important in the case of very compact binaries of the type recently detected at TeV $\gamma$-ray energies. Only these calculations can be applied for 
the case of the simulation of the full problem of the anisotropic cascade initiated by primary leptons or $\gamma$-rays in the radiation field of the massive star since in such a case parts of the cascade can also develop very close to the surface of the massive star.

Note, that these early calculations of the optical depths can be simply scaled for the massive stars with other parameters, i.e. radius and surface temperature. 
Let us denote the parameters of the reference calculations by stars and the parameters of the optical depths for other stars by $N$. Then, this simple scaling formula can be described as,
\begin{eqnarray}
\tau (E_{\gamma}^{\rm N} = 
E_\gamma^\star/S_{\rm T}, T^{\rm N}, R^{\rm N}, D, \alpha ) = S_{\rm T}^3S_{\rm R}\tau (E_\gamma^\star,T^\star,R^\star,D,\alpha),
\end{eqnarray}
\noindent
where $S_{\rm T} = T^{\rm N}/T^\star$, $S_{\rm R} = R^{\rm N}/R^\star$, distance from the star , D, is measured in the stellar units, and $\alpha$ is the injection angle.
Therefore {\it there is no need for separate calculations of the optical depths for the stars with other parameters}. Such example calculations which may serve as the base for the scaling procedure according to Eq.~1 has been shown in Fig.~\ref{fig1} (see also Bednarek 2000).

The optical depths inside most of the massive binary systems mentioned above, for the location of the $\gamma$-ray source at the distance of the compact object from the massive star, can reach  values above ten. The only exception is the system PSR 1259-63/SS2883. However, as we have shown in Sierpowska-Bartosik \& Bednarek~(2008), it is likely that electrons also in this binary system can be accelerated much closer to the surface of the massive star than the distance of the periastron passage. At such distances the optical depths for $\gamma$-rays can be larger than unity. 
Therefore, we conclude that when calculating the $\gamma$-ray spectra from all these binary systems, the cascade processes has to be taken into account.

\section{Cascade in the radiation of massive star}

The acceleration of electrons injected somewhere inside the binary system occurs
in strong radiation field of the nearby massive star and also in its magnetic field.
The maximum energies of electrons depend on the efficiency of
their energy losses on these two processes, on the efficiency of acceleration process,
and on the escape conditions from acceleration region. Moreover, in some situations also the radiation produced by the compact object itself can serve as a target for electrons. Therefore, the spectrum and maximum energies of electrons, depend on the conditions which are not usually well known and some simplified assumptions are necessary.
We limit our considerations only to the case when the dominant energy loss mechanism is provided by the synchrotron and IC energy losses (the radiation and magnetic field of the massive star can be uniquely defined).

In order to consider the radiation processes of accelerated electrons, someone
usually assumes that electrons are injected mono-energetically or with the power law spectrum.
Electrons produce first generation of $\gamma$-rays by IC scattering of stellar radiation with specific angular distribution, since they are injected into anisotropic 
radiation field (the stellar radiation arrives from specific directions in respect to electron). Moreover, secondary $\gamma$-rays see also anisotropic radiation field of the massive star. So then, their escape/absorption conditions strongly depend on their direction in
respect to the massive star. 
Therefore, the whole process of $\gamma$-ray production in such scenario becomes very complicated even without including secondary processes such as synchrotron losses of secondary leptons.
Such cascade process can be considered in the three limiting cases:

\begin{enumerate}

\item One-dimensional linear cascade.

\item Three dimensional cascade with isotropization of secondary leptons.

\item Three dimensional cascade with leptons following the magnetic field.

\end{enumerate}

All these cases are approximations of the case realized in practice and they might give reasonable  approximation in specific parts of the binary system.
In the case (1), secondary leptons, produced as a result of absorption of $\gamma$-rays in 
the stellar radiation, propagate linearly  in the direction of the parent $\gamma$-ray photon.
This is the simplest possible case but also the least realistic for $\gamma$-rays injected relatively close to the massive star ($<10 R_\star$). This is due to the fact that the Larmor radius of secondary leptons in a typical magnetic field of the massive star is
significantly lower than the characteristic interaction length on the IC scattering of TeV leptons in 
the radiation field of the massive star. As an example, let as consider the case of the 
massive star with the surface magnetic field of the order of 300 G. At the characteristic 
distance of the TeV $\gamma$-ray source inside massive binary system (equal to $R\sim 5 R_\star$) the magnetic field strength is $\sim 3$ G. The Larmor radius of 1 TeV leptons is then equal to $R_{\rm L}\approx  10^9$ cm. On the other hand, the 
characteristic distance scale for IC scattering in the Thomson regime in the radiation of the massive star with surface temperature $3\times 10^4$ K and above mentioned distance, $R = 5R^\star$,  is $\lambda\approx   (n_\star\sigma_{\rm T})^{-1}\approx 10^{11}$ cm. This is clearly larger than $R_{\rm L}$ which means that leptons are isotropized or driven by the magnetic field before interacting with stellar photons. Note, that
IC scattering of secondary leptons with TeV energies occurs in fact in the KN regime and the difference between $\lambda$ and $R_{\rm L}$ is larger for distances closer to the massive star at which the compact object is located. 
These two additional effects strengthen the conclusion that inside the binary system
secondary leptons can not follow the direction of the parent $\gamma$-ray photon and their
trajectories are significantly re-directed before the IC interaction with the stellar photon.

Therefore, we consider only the cascade scenarios according to (2) and (3). 
The difference between these two cases depends on the domination of the main component of 
the magnetic field around the massive star. In the earlier approach (2), it has been assumed that the random component is enough important to isotropize secondary leptons produced in the cascade process. This occurs when the Larmor radius of secondary leptons is much smaller than the characteristic ICS length. Also the synchrotron energy losses are neglected in respect to IC energy losses  (for discussion of these conditions see Sect.~2 in Bednarek~1997)
In the recent calculations (3), we follow the paths of secondary leptons in  the magnetic field with the dipole structure and also subtract their energy losses on synchrotron process
(see details in Sierpowska \& Bednarek~(2005).

\section{Scenarios for $\gamma$-ray production inside binary}

Three geometrical scenarios for the acceleration of electrons has been considered up to now,
i.e. injection of particles: (a) along the shock structure which appears due to the presence of  energetic pulsar or by pulsar itself; (b) along the jet from the accretion disk around the compact object; (c) along the shock structure created in collisions of stellar winds.  

\subsection{Energetic pulsar close to massive star}

In this case leptons can be directly accelerated in the pulsar magnetosphere or at the shock (in classical shock acceleration scenario) which appears as a result of collision between the pulsar and stellar winds (see Fig.~2). If the pulsar and stellar winds and spherical then the shock structure is axially symmetric and easy to consider. However, in the case of the Be type stars, the winds are composed of two parts: the equatorial low velocity but dense wind and the polar fast and rare wind. Then, the shock can reach very complicated structure with some parts very close to the  massive star (see Fig~2). In this last case, which we consider for the binary PSR 1259-63/SS2883 (Bednarek \& Sierpowska~2008), leptons can be injected into the binary system even at a few stellar radii in spite of the periastron passage of the pulsar at the distance of $23 R_\star$ from the star. 

\begin{figure*}[t]
\vskip 6.truecm
\includegraphics{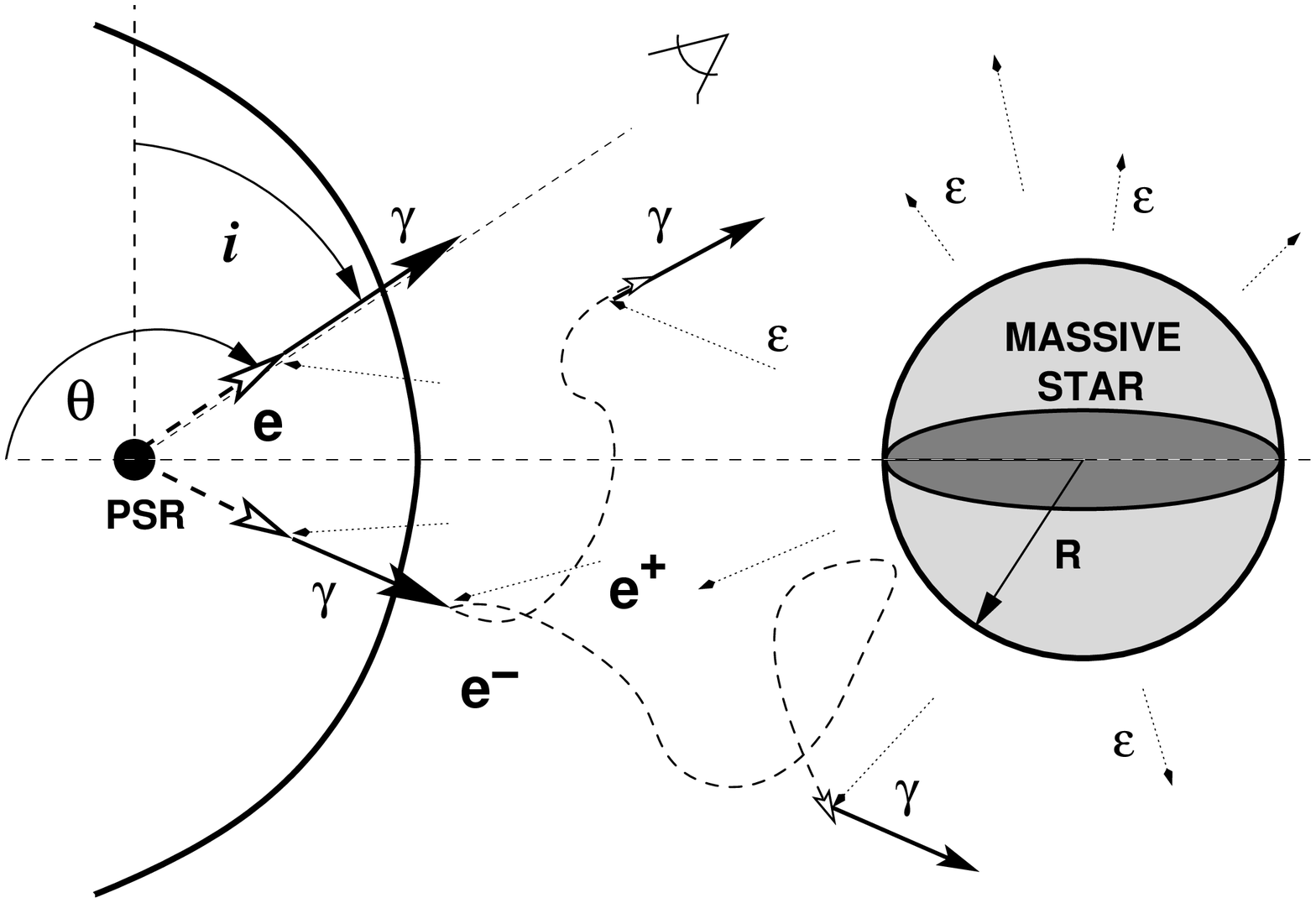}
\includegraphics{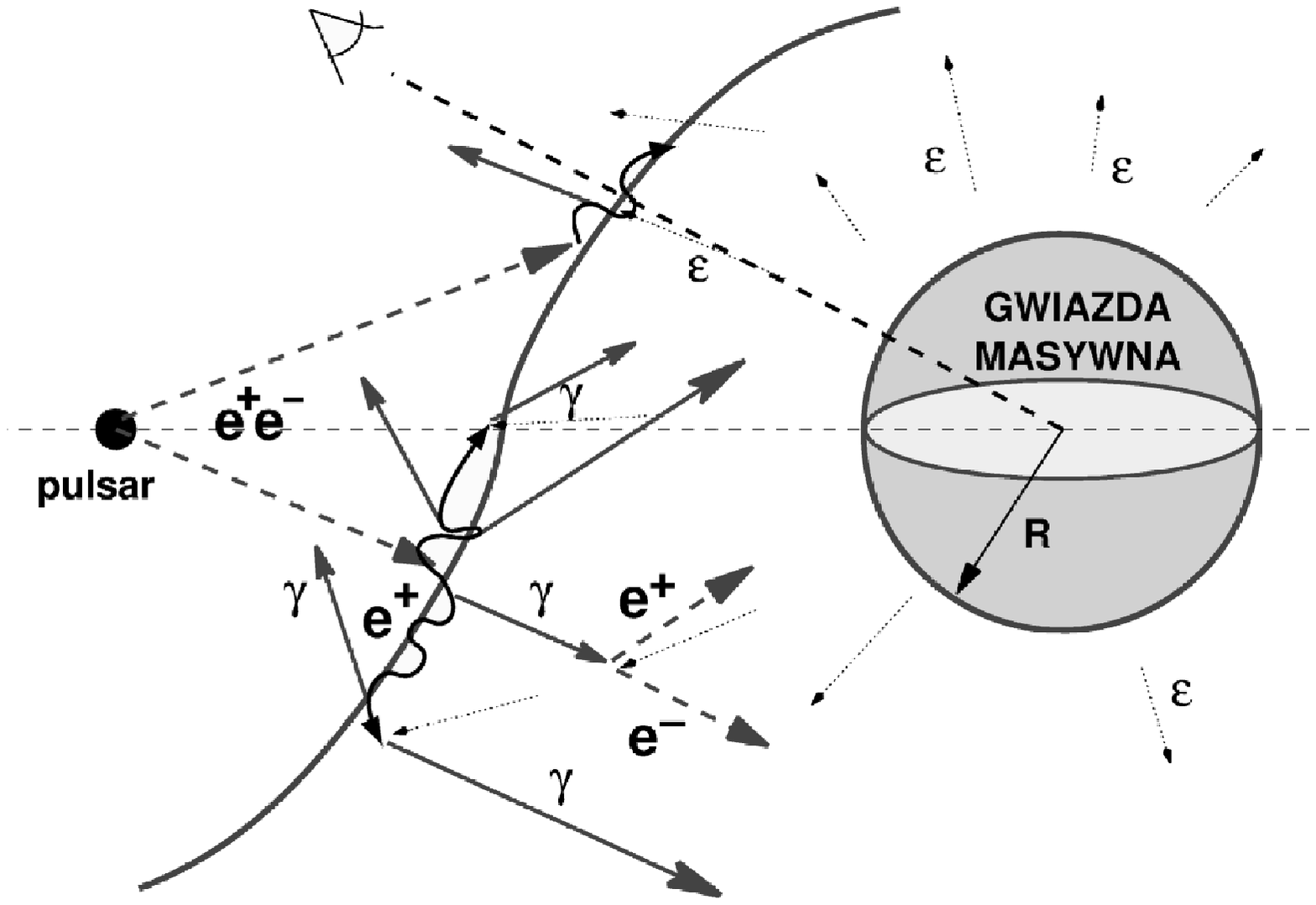}
\caption{The scenarios for the $\gamma$-ray production in the IC $e^\pm$ 
pair cascade process initiated by leptons from the pulsar or accelerated at the 
shock formed in collisions of pulsar and stellar winds in the cases of
spherical winds (left) and a pulsar wind with a non-spherical wind of massive star (right).} 
\label{fig2}
\end{figure*}

The cascade $\gamma$-ray spectra and light curves in the case of the 
pulsar inside the spherical stellar wind has been shown for the example case of the 
binary system with the Cyg X-3 parameters in Sierpowska \& Bednarek~(2005). We concluded 
that the distribution of the cascade $\gamma$-ray spectra on the sky shows interesting structures which are determined by the dipole structure of the magnetic field of the massive star and the injection directions of primary particles. The case of the pulsar wind interacting with the non-spherical stellar wind has been discussed with the application to the Be binary system PSR 1259-63/SS2883 (see Sierpowska-Bartosik \& Bednarek~2008). We show
that even in the case of this relatively extended binary system the cascade calculations should be applied for some parts of the shock structure which can even appear at the distance  
of a few stellar radii from the surface of the Be star. We calculated the $\gamma$-ray light curve close to the periastron passage of the neutron star. It has general features similar to those observed by the HESS telescopes (Aharonian et al.~2005a).

The detailed studies of the spectral and time characteristics of the TeV $\gamma$-ray binary system LS 5039 under the assumption of the presence of energetic pulsar have been recently performed by Sierpowska-Bartosik \& Torres~(2007, 2008).

\subsection{Jet from an accretion disk around compact object}

These type of objects, called microquasars, can accelerate electrons
along the straight jet. The process of acceleration is not well known. 
Modelling of such sources needs to fix various parameters which describe acceleration process such as: acceleration efficiency along the jet and along the orbit of the compact object,  maximum energies of electrons, jet direction in respect to the plane of the binary.
Therefore, the final results are strongly model dependent. However, other parameters, such as the orbit of  the binary, the massive companion, location of the observer are usually quite well known, making these systems better defined than e.g. in the case of $\gamma$-ray production in AGNs jets. 

The calculations of $\gamma$-ray spectra from the IC $e^\pm$ pair cascades ini\-tia\-ted by
primary electrons inside the jet has been done with application to three massive binary systems: LS 5039 (Bednarek~2006a), LSI +61$^{\rm o}$ 303 (Bednarek~2006b), and
Cyg X-1 (Bednarek \& Giovannelli~2007). It has been shown that in general the maximum of the TeV $\gamma$-ray emission from these massive binaries should appear close to the phases when the compact object is in front of the massive star (in the case of circular orbit).
If the orbit of the compact object is very extended, than the phase of the periastron passage, in respect to the phase of the observer,  can significantly change the simple pattern expected for the circular orbit. Moreover, the IC $e^\pm$ pair cascade model for these binary systems predicts unticorrelation of the TeV $\gamma$-ray emission in respect to GeV emission (see also early calculations in Bednarek~2000). This prediction should be tested by the future observations with the LAT telescope on the board of GLAST.

\begin{figure*}[t]
\vskip 6.truecm
\includegraphics{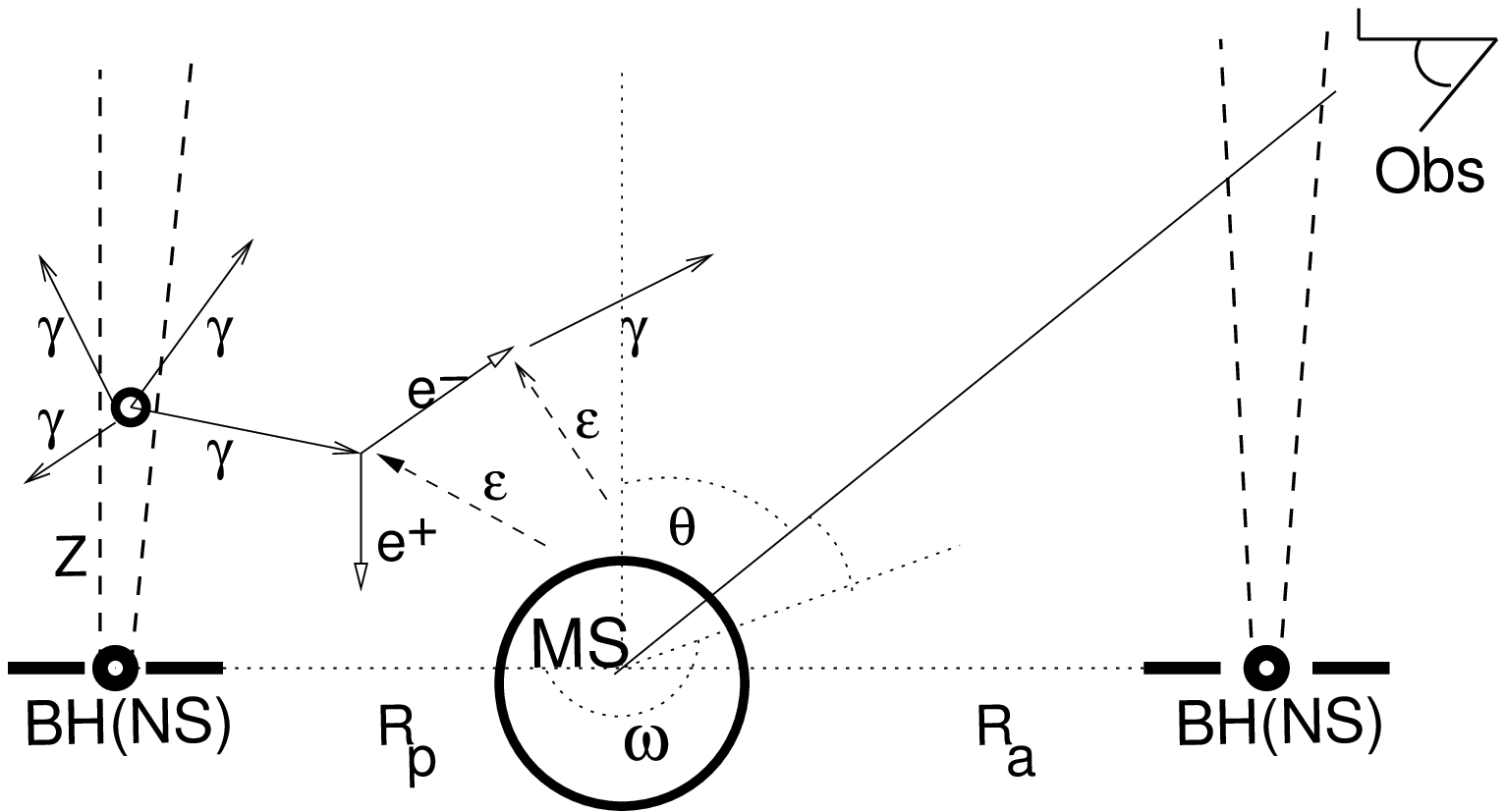}
\includegraphics{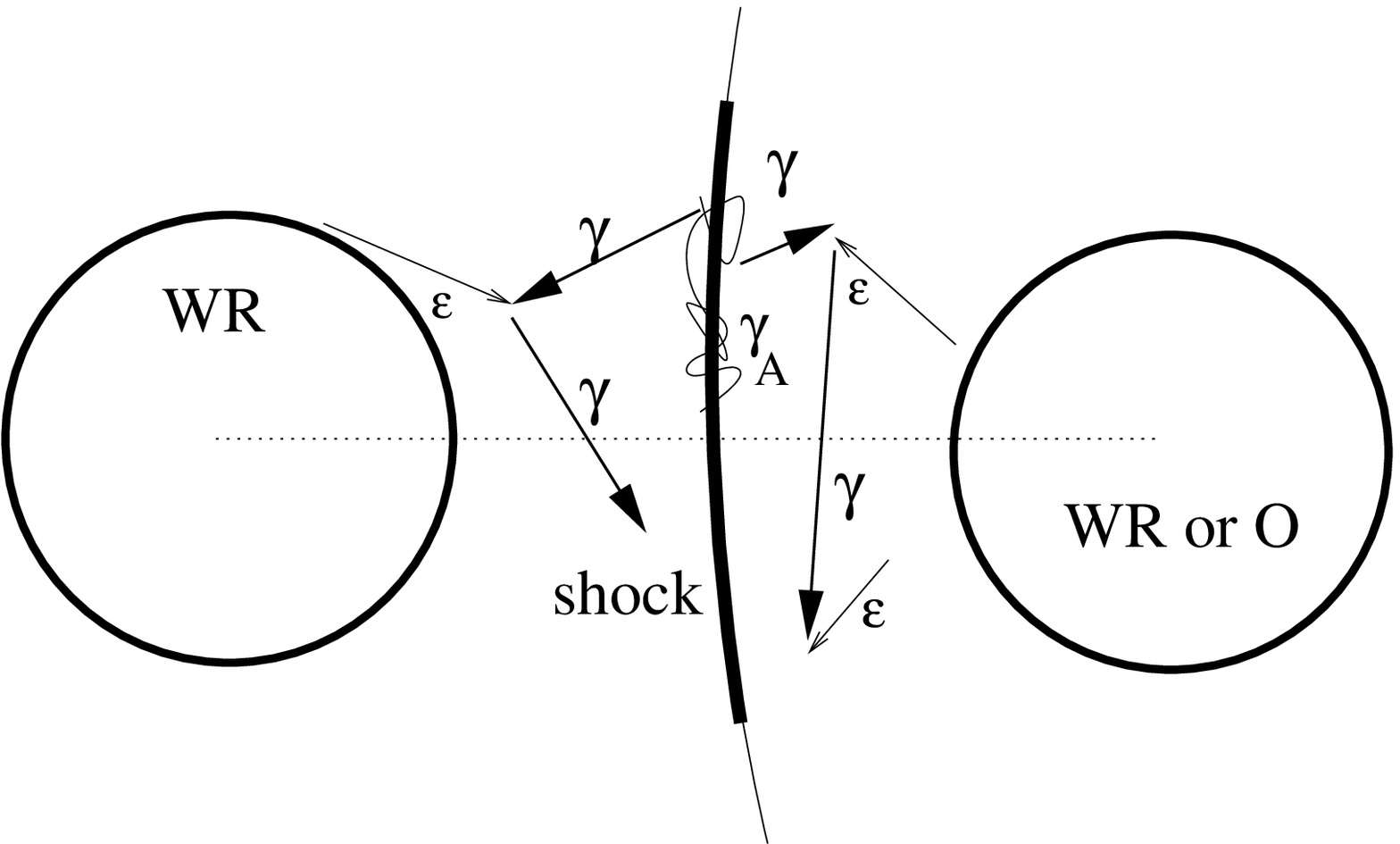}
\caption{The scenarios for $\gamma$-ray production in the IC $e^\pm$ pair cascade initiated by leptons accelerated along the jet from the central parts of the accretion disk around compact object nearby to the massive star (left figure) and initiated by leptons and primary $\gamma$-rays from decay of pions produced in hadronic collisions of accelerated nuclei with the matter of the stellar winds from two massive star binary (right).} 
\label{fig3}
\end{figure*}
\subsection{Collision of stellar winds}

Particles can be also accelerated at the shock wave formed in collisions of fast winds from two
massive stars in compact binary system. As an example, we discussed (Bednarek~2005) the most massive binary of two WR stars, WR 20a, in the open cluster Westerlund 2. However, in this case, due to extremely strong radiation field, electron acceleration is saturated by huge IC energy losses at energies which prevent TeV $\gamma$-ray production. Therefore, we consider  acceleration of nuclei to characteristic Lorentz factors $\sim 10^{5-7}$, which interacting with 
the matter  of the stellar winds produce pions. These pions decay to $\sim$TeV $\gamma$-rays and leptons initiating IC $e^\pm$ pair cascade in the stellar radiation. In the case of WR 20a, we were able to normalize the GeV emission produced in the cascade to the flux of the EGRET sources 
observed in the direction of WR 20a. On this base, the TeV $\gamma$-ray flux has been predicted
on the level which should be detectable by the southern hemisphere Cherenkov telescopes.
In fact, the TeV $\gamma$-ray source has been discovered from the direction of WR 20a (Aharonian et al.~2007), but it does not necessarily can be linked to WR 20a due to its significant extend.


\bigskip
\bigskip
\noindent {\bf DISCUSSION}

\bigskip
\noindent {\bf JIM BEALL:} Can you say again of the maximum gamma of the hadrons? 

\bigskip
\noindent {\bf WŁODEK BEDNAREK:} The maximum Lorentz factors of 
hadronos accelerated in the wind collision model of two massive stars is of the order of
a few $10^6$ (Bednarek~2005). 

\end{document}